\documentclass{optica-article}

\usepackage[table]{xcolor}
\usepackage{colortbl}
\makeatletter
\def\@journalname{arXiv preprint}
\makeatother

\articletype{Research Article}

\usepackage{amsmath}
\usepackage{subcaption}

\begin{document}

\title{Introducing an Extensible Open-Source Toolkit Suite for Studying Second Harmonic Generation: A Case Study of Depleted Pulsed Gaussian Wave SHG}

\author{Mostafa M. Rezaee,\authormark{1,2}
        Mohammad Sabaeian,\authormark{2}
        Alireza Motazedian,\authormark{3,2}
        Fatemeh Sedaghat Jalil-Abadi,\authormark{4,2} and
        Mohammad Ghadri\authormark{5,*}}
\address{
    \authormark{1}Data Science Program, Bowling Green State University, Bowling Green, OH, USA\\
\authormark{2}Department of Physics, Shahid Chamran University of Ahvaz, Ahvaz, Khuzestan, Iran\\
\authormark{3}Department of Physics, University of New Hampshire, NH, USA\\
\authormark{4}Department of Energy Engineering and Physics, Amirkabir University of Technology, Tehran, Iran\\
\authormark{5}MIAE Department, Concordia University, Montreal, QC, Canada}

\email{\authormark{*}mohammad.ghadri@mail.concordia.ca}

\begin{abstract*}
Second Harmonic Generation (SHG) in nonlinear crystals has been extensively investigated, but most existing models still rely on  simplifying assumptions. In realistic settings, thermal effects introduce complications that are difficult to capture analytically because the governing equations are highly coupled and nonlinear. Direct experimental characterization is also limited, since studying thermal effects would require spatiotemporal temperature data at every point in the crystal, which is not experimentally accessible.

To address these limitations, we have developed a SHG Computational Toolkit Suite, a coordinated collection of independent modeling toolkits that cover different SHG scenarios under various physical conditions. Each toolkit focuses on a particular configuration or coupling mechanism, while the suite as a whole provides well-documented numerical implementations, reproducible workflows, and illustrative examples. 

Together, this article and the Toolkit Suite provide a coherent infrastructure for computational studies of SHG. It enables researchers to replicate, adapt, and extend our methods without duplicating foundational development efforts, thereby accelerating SHG research and promoting reproducibility.

\end{abstract*}


%%%%%%%%%%%%%%%%%%%%%%%%%%  body  %%%%%%%%%%%%%%%%%%%%%%%%%%

\section{Introduction}
\noindent
Second Harmonic Generation (SHG) remains an important area of modern optics research \cite{PhysRevB.104.L161202, Briggs:24, nano14080662, aghigh2023second, zu2024optical, liu2019phase, AKRAMOV2024129827}. However, accurate modeling of SHG becomes increasingly challenging in regimes involving thermal effects, phase mismatch under different beam configurations, where the underlying optical and thermal processes are strongly coupled and nonlinear.

One major challenge is the need to understand and manage dissipated heat within the nonlinear crystal, which in principle requires access to the spatiotemporal temperature distribution during the SHG process \cite{sabaeian2008analytical, rezaee2015thermally}. In practice, such internal temperature measurements are not feasible, making direct experimental characterization impractical \cite{aghigh2023second}. Analytical approaches can provide valuable insight \cite{zu2024optical, zu2022analytical, li2024highly}, but they often rely on simplifying assumptions that limit their applicability under realistic operating conditions \cite{Zhao:20}. These constraints motivate the need for flexible computational approaches capable of capturing diverse SHG phenomena, including thermal dynamics \cite{sabaeian2015temperature, Sabaeian_2026_CWG-Ideal}.

Computational techniques offer a practical alternative by enabling numerical modeling of SHG’s nonlinear and interconnected processes without relying exclusively on restrictive assumptions or inaccessible experimental measurements \cite{zu2022analytical, Carnio:20, sabaeian2015tempdist, Rezaee2025ThermalKTP, sabaeian2014heat}. They also make it possible to perform extensive parameter sweeps and simulations within reasonable computation times \cite{Chou:98}.

Numerous computational approaches to SHG have been proposed \cite{Szarvas:18, Carnio:20, saito2016numerical}, including our own previous works \cite{sabaeian2015tempdist, rezaee2015complete, rezaee2015thermally, sabaeian2015temperature, sabaeian2014pulsed, sabaeian2014heat}. However, the gap is not the absence of computational SHG models, but the lack of a fully accessible, well-documented, and extensible resource that consolidates validated methods across diverse SHG configurations. Existing tools remain limited in their ability to support reproducible workflows, modular extension, and adaptation to more complex nonlinear and thermo-optical scenarios \cite{boudjema2020pyshs, paez2024cuda}.

To address this gap, we developed a Computational \href{https://github.com/Second-Harmonic-Generation}{SHG Toolkit Suite}, a publicly accessible and well-documented GitHub Organization that provides numerical simulation and data analysis capabilities for advanced SHG investigations. The suite provides: (1) a modular architecture in which independent toolkits can be combined and adapted for specific scenarios, (2) comprehensive documentation of both the underlying physics and numerical implementation details, (3) validated implementations for depleted regimes, thermal effects, and pulsed configurations that are ready for research use, and (4) full source-code accessibility that enables researchers to modify, extend, and contribute to the methods. The suite covers continuous-wave and pulsed SHG in Gaussian and Bessel--Gaussian beams, with and without thermal and phase-mismatch effects. 

Taken together, these repositories form a coordinated toolkit suite in which common numerical and physical components can be reused, recombined, and adapted to different beam types, crystal configurations, and coupling mechanisms. The work provides researchers with a practical foundation for building upon validated methods without reconstructing the computational infrastructure from scratch \cite{van2024encore, dogucu2024reproducibility}.

This contribution also supports the broader movement toward computationally supported physics education \cite{fredly2024computational, odden2023using, weller2022development, phillips2023physicality}. The SHG Toolkit Suite provides documented, executable examples that allow users to examine governing equations, run simulations, and modify parameters. Its modular architecture facilitates future extensions for advanced SHG scenarios involving heat generation, phase mismatch, and strongly coupled nonlinear dynamics \cite{rezaee2015thermally, Rezaee_2026_PWG-PhM}.

To demonstrate this extensibility in a realistic research-grade setting, we present a case study in which the suite is extended to model pulsed Gaussian second-harmonic waves generated under a Type II configuration. The model employs the Finite Difference Method (FDM), assumes ideal coupling of the ordinary and extraordinary fundamental beams in KTP, and neglects thermal absorption. Cylindrical coordinates are used to exploit the azimuthal symmetry of the pump profile, and three coupled equations describe the evolution of the two fundamental beams and the generated second-harmonic wave. Although G. D. Boyd et al.\ previously treated these equations under simplifying assumptions \cite{boyd1968parametric}, the present FDM-based implementation enables a broader and more accurate numerical exploration of SHG configurations and provides a concrete example of how the toolkit suite can be adapted for new research problems.

\section{Theory}
\noindent
The pulsed SHG equations derive from Maxwell's equations. As a starting point, we consider the Helmholtz equation in steady state for a dispersive and nonlinear medium, written for a monochromatic polarized field as follows \cite{Boyd2003Nonlinear_19}:

\begin{equation}
    \nabla^{2} \vec{E}_{n}(r, z, t)+\frac{\omega_{n}^{2}}{c^{2}} \varepsilon^{(l)}\left(\omega_{n}\right) \cdot \vec{E}(r, z, t)=-\frac{\omega_{n}^{2}}{\varepsilon_{0} c^{2}} \vec{P}_{n}^{N L}(r, z, t)
    \label{eq:1}
\end{equation}

\noindent
Here $\vec{E}_{n}(r, z, t)$ is the electric field, $\vec{P}_{n}^{N L}(r, z, t)$ is the nonlinear polarization, and $c$ and $\varepsilon^{(l)}$ are the speed of light and the dielectric tensor, respectively. To include absorption in the medium, the dielectric tensor is written as

\begin{equation}
    \varepsilon^{l}=\varepsilon_{r}^{l}+i \varepsilon_{i m}^{l}
    \label{eq:2}
\end{equation}

The field and nonlinear polarization can be written as \cite{feng2012efficient_20}:

\begin{equation}
    P_{n}^{N L}(r, z, t)=P_{n}^{N L}(r, z) e^{i k_{n} z-i \omega_{n} t}+C . C
    \label{eq:3}
\end{equation}

\begin{equation}
    E_{n}(r, z, t)=E_{n}(r, z) e^{i k_{n} z-i \omega_{n} t}+C . C
    \label{eq:4}
\end{equation}

\noindent
Here C.C denotes the complex conjugate, $\omega$ is the angular frequency, and $k$ is the wave number. Due to azimuthal symmetry, only the $r$ and $z$ coordinates are retained, and the Laplacian is taken only in these directions, i.e.\ $\nabla^{2}=\nabla_{r}^{2}+\nabla_{z}^{2}$.
Following standard derivation procedures, the following expression is obtained:

\begin{equation}
    \frac{d \vec{E}_{n}(r, z, t)}{d z}-\frac{i}{2 k} \nabla_{r}^{2} \vec{E}_{n}(r, z, t)+\frac{\gamma}{2} \vec{E}_{n}(r, z, t)=\frac{i \omega_{n}}{2 n_{n} \varepsilon_{0} c} \vec{P}_{n}^{N}(r, z, t)
    \label{eq:5}
\end{equation}

\noindent
Here, $\varepsilon_{\mathrm{o}}=8.85 \times 10^{-12} \,\mathrm{C}^{2} / \mathrm{N m}^{2}$ is the vacuum permittivity and $\gamma=\varepsilon_{i m}\frac{\omega}{n c}$ is the absorption coefficient.

For pulsed beams, the field envelope varies slowly in time compared to the optical oscillation period. To account for this, we generalize the envelope from $E_n(r,z)$ to $E_n(r,z,t)$, where the time dependence captures the pulse shape. Applying the slowly-varying envelope approximation (SVEA) in time introduces an additional term $(n/c)\partial E/\partial t$ in the propagation equations.

Including temporal evolution, the nonlinear polarization sources for the fundamental beams and second harmonic beam are introduced as follows \cite{sabaeian2010investigation_17,Boyd2003Nonlinear_19}:

\begin{equation}
    P_{1}^{N L}=4 \varepsilon_{0} d_{e f f} E_{3}(r, z) E_{2}^{*}(r, z) e^{i\left(k_{2}-k_{3}\right) z} e^{-i \omega_{1} t}
    \label{eq:6}
\end{equation}

For $\omega_1 = \omega$, and

\begin{equation}
    P_{2}^{N L}=4 \varepsilon_{0} d_{e f f} E_{3}(r, z) E_{1}^{*}(r, z) e^{i\left(k_{1}-k_{3}\right) z} e^{-i \omega_{2} t}
    \label{eq:7}
\end{equation}

For $\omega_2 = \omega$, and for $\omega_3 = 2\omega$

\begin{equation}
    P_{3}^{N L}=4 \varepsilon_{0} d_{e f f} E_{1}(r, z) E_{2}(r, z) e^{i\left(k_{1}+k_{2}\right) z} e^{-i \omega_{3} t}
    \label{eq:8}
\end{equation}

A set of coupled equations is written as below

\begin{equation}
    \frac{n_{1}}{c} \frac{d E_{1}}{d t}+\frac{d E_{1}}{d z}-\frac{i c}{2 n_{1} \omega} \nabla_{r}^{2} E_{1}+\frac{\gamma_{1}}{2} E_{1}=\frac{2 i \omega}{n_{1} c} d_{eff} E_{2}^{*} E_{3} e^{-i \Delta k z}
    \label{eq:9}
\end{equation}

\begin{equation}
    \frac{n_{2}}{c} \frac{d E_{2}}{d t}+\frac{d E_{2}}{d z}-\frac{i c}{2 n_{2} \omega} \nabla_{r}^{2} E_{2}+\frac{\gamma_{2}}{2} E_{2}=\frac{2 i \omega}{n_{2} c} d_{eff} E_{1}^{*} E_{3} e^{-i \Delta k z}
    \label{eq:10}
\end{equation}

\begin{equation}
    \frac{n_{3}}{c} \frac{d E_{3}}{d t}+\frac{d E_{3}}{d z}-\frac{i c}{4 n_{3} \omega} \nabla_{r}^{2} E_{3}+\frac{\gamma_{3}}{2} E_{3}=\frac{4 i \omega}{n_{3} c} d_{eff} E_{1} E_{2} e^{i \Delta k z}
    \label{eq:11}
\end{equation}

\noindent
The factor of 4 in Eq.~\eqref{eq:11} (compared to the factor of 2 in Eqs.~\eqref{eq:9}--\eqref{eq:10}) arises from the doubled frequency of the second harmonic wave ($\omega_3 = 2\omega$), which affects both the energy normalization and the coupling strength in the nonlinear interaction.

Here $d_{eff}=2 \chi^{(2)}$ is the effective nonlinear coefficient \cite{Boyd2003Nonlinear_19}. In type II SHG, $\omega_1=\omega_2=\omega$ and $\omega_3=\omega_1+\omega_2=2\omega$. For perfect phase matching, the wave vector mismatch must vanish, i.e., $\Delta k={k_1}+{k_2}-{k_3}=0$, which corresponds to $\Delta \phi=\Delta k z=0$. Alternatively, if phase mismatch effects are neglected, we set $\Delta \phi=0$ directly. Phase matching can be achieved if $n^{\omega, o} \omega+n^{\omega, e} \omega=n^{2 \omega, e} 2 \omega$; in the above expressions $n_{1}=n^{\omega, o}, n_{2}=n^{\omega, e}, n_{3}=n^{2 \omega, e}$ \cite{sabaeian2010investigation_17}. With the refractive indices listed in Table~\ref{tab:1}, this condition is satisfied: $n_1 + n_2 = 1.8296 + 1.7466 = 3.5762 = 2 \times 1.7881 = 2n_3$.

Having derived all the equations needed to generate the second harmonic wave, we now introduce dimensionless quantities to reduce numerical errors. Specifically, we define variables as ratios of the generated wave intensity to the initial wave intensity:

\begin{equation}
    \psi_{1}=\frac{E_{1}}{\sqrt{P_{1} / 2 n_{1} c \varepsilon_{0} \pi \omega_{f}^{2}}} \;\Rightarrow\; E_{1}=\sqrt{\frac{P_{1}}{2 n_{1} c \varepsilon_{0} \pi \omega_{f}^{2}}} \psi_{1} \;\Rightarrow\; \eta_{1}= \left|\psi_{1}\right|^{2}=\frac{I_{1}}{I_{1}(0)}
    \label{eq:12}
\end{equation}

\begin{equation}
    \psi_{2}=\frac{E_{2}}{\sqrt{P_{2} / 2 n_{2} c \varepsilon_{0} \pi \omega_{f}^{2}}} \;\Rightarrow\; E_{2}=\sqrt{\frac{P_{2}}{2 n_{2} c \varepsilon_{0} \pi \omega_{f}^{2}}} \psi_{2} \;\Rightarrow\; \eta_{2}=\left|\psi_{2}\right|^{2}=\frac{I_{2}}{I_{2}(0)}
    \label{equ:13}
\end{equation}

\begin{equation}
    \psi_{3}=\frac{E_{3}}{\sqrt{P_{3} / 2 n_{3} c \varepsilon_{0} \pi \omega_{f}^{2}}} \;\Rightarrow\; E_{3}=\sqrt{\frac{P_{3}}{2 n_{3} c \varepsilon_{0} \pi \omega_{f}^{2}}} \psi_{3} \;\Rightarrow\; \eta_{3}=\left|\psi_{3}\right|^{2}=\frac{I_{3}}{I_{1}(0)+I_{2}(0)}
    \label{eq:14}
\end{equation}

In which the quantity $P_i$ and $\eta_{i}$ with $i =$ 1, 2, and 3, give the power and intensity efficiency, respectively, and $\omega_f$ denotes the fundamental beam spot size. As the fundamental waves have the same frequency, their power with orthogonal polarization is equal, as well. Hence, in the $z=0$ plane, the power of the fundamental wave $P_1 = P_2 = P$ and via the SHG approach the power of the final wave is equal to $P_3 = 2P$. Replacing the change of variable from Eqs.~\eqref{eq:12}--\eqref{eq:14} into Eqs.~\eqref{eq:9}--\eqref{eq:11}, the three type II SHG equations are obtained as below:

\begin{equation}
    \frac{n_{1}}{c} \frac{d \psi_{1}}{d t}+\frac{d \psi_{1}}{d z}-\frac{i c}{2 n_{1} \omega} \frac{1}{r} \frac{d \psi_{1}}{d r}-\frac{i c}{2 n_{1} \omega} \frac{d^{2} \psi_{1}}{d r^{2}}+\frac{\gamma_{1}}{2} \psi_{1}=\frac{i}{L} \psi_{2}^{*} \psi_{3} e^{-i \Delta \phi}
    \label{eq:15}
\end{equation}

\begin{equation}
    \frac{n_{2}}{c} \frac{d \psi_{2}}{d t}+\frac{d \psi_{2}}{d z}-\frac{i c}{2 n_{2} \omega} \frac{1}{r} \frac{d \psi_{2}}{d r}-\frac{i c}{2 n_{2} \omega} \frac{d^{2} \psi_{2}}{d r^{2}}+\frac{\gamma_{2}}{2} \psi_{2}=\frac{i}{L} \psi_{1}^{*} \psi_{3} e^{-i \Delta \phi}
    \label{eq:16}
\end{equation}

\begin{equation}
    \frac{n_{3}}{c} \frac{d \psi_{3}}{d t}+\frac{d \psi_{3}}{d z}-\frac{i c}{4 n_{3} \omega} \frac{1}{r} \frac{d \psi_{3}}{d r}-\frac{i c}{4 n_{3} \omega} \frac{d^{2} \psi_{3}}{d r^{2}}+\frac{\gamma_{3}}{2} \psi_{3}=\frac{i}{L} \psi_{1} \psi_{2} e^{i \Delta \phi}
    \label{eq:17}
\end{equation}

The interaction length L is defined by

\begin{equation}
    L=\left(\frac{n_{1} n_{2} n_{3} c^{3} \varepsilon_{0} \pi \omega_{f}^{2}}{4 P \omega^{2} d_{e f f}^{2}}\right)^{\frac{1}{2}}
    \label{eq:18}
\end{equation}

This quantity merits further investigation. It represents a characteristic length over which the nonlinear interaction is strongest. In other words, this quantity provides a clear measure of how the effective parameters of a nonlinear laser influence the interaction. The nonlinear effective coefficient $d_{eff}$ and the refractive indices are intrinsic properties of a given nonlinear crystal and are therefore fixed for that crystal, whereas the fundamental wave power and the spot size at the entrance face can be adjusted and thus affect the SHW efficiency.

As mentioned earlier, Gaussian beams are the focus of this study. Thus, we presume a Gaussian beam for the laser source or fundamental beams as boundary conditions at the crystal input plane, where

\begin{equation}
    \psi_{1}(r,0) = \psi_{2}(r,0) = \exp\!\left(-\frac{r^{2}}{\omega_{f}^{2}}\right), \qquad \psi_{3}(r,0) = 0.
\end{equation}

Time-dependent boundary conditions at $z=0$ for fundamental and second harmonic beams should be taken into account as

\begin{equation}
    \psi_{1}(t, r, z=0)=\exp \left[-\left(t / t_{p}\right)^{2}\right] \times \exp \left(-r^{2} / \omega_{f}^{2}\right)
    \label{eq:19}
\end{equation}

\begin{equation}
    \psi_{2}(t, r, z=0)=\exp \left[-\left(t / t_{p}\right)^{2}\right] \times \exp \left(-r^{2} / \omega_{f}^{2}\right)
    \label{eq:20}
\end{equation}

\begin{equation}
    \psi_{3}(t, r, z=0)=0
    \label{eq:21}
\end{equation}

where $t_p$ is the pulse duration. At the beam axis ($r=0$), the axial symmetry of the problem requires $\partial \psi / \partial r = 0$, which is implemented numerically by setting $\psi(r=0) = \psi(r=\Delta r)$.

The initial conditions are specified as follows: (1) For $t < 0$, all fields are zero throughout the crystal, representing the state before the pulse arrives. (2) At $t = 0$, the fields inside the crystal ($z > 0$) are initialized to zero, while the boundary conditions at the input face ($z = 0$) are given by Eqs.~\eqref{eq:19}--\eqref{eq:21}. (3) The pulse is ``turned on'' through the time-dependent boundary conditions at $z = 0$: the Gaussian temporal profile $\exp[-(t/t_p)^2]$ ensures that the pulse amplitude is negligible for $t \ll -t_p$, grows smoothly as time progresses, reaches its maximum at $t = 0$, and then decays. This formulation naturally handles the pulse onset without requiring explicit switching mechanisms, as the exponential decay ensures the pulse amplitude is effectively zero before the simulation begins.

\section{Results and discussion}
\noindent
By applying the toolkit to a depleted pulsed Gaussian wave SHG configuration, we illustrate how researchers can leverage the suite's modular architecture, comprehensive documentation, validated implementations, and accessibility to address specific problems without reconstructing computational infrastructure from scratch. This example validates the toolkit's functionality while providing a concrete template for adaptation to other SHG scenarios.

The coupled field equations (Eqs.~\eqref{eq:15}--\eqref{eq:17}), together with the boundary and initial conditions described above, are solved numerically using a self-developed FORTRAN code running under the Linux Ubuntu operating system. The equations are discretized in cylindrical coordinates using the finite difference method (FDM). Backward FDM is used for the temporal derivatives in Eqs.~\eqref{eq:15}--\eqref{eq:17}, forward FDM for the spatial derivatives along the crystal axis, and central FDM for the radial derivatives. The numerical methodology employed here follows the same validated approach used in our previous studies \cite{sabaeian2014pulsed, rezaee2015complete, Rezaee_2026_CWG_Heat}, where the FDM-based solution of coupled SHG equations was validated against analytical solutions and experimental data for similar configurations.

This case study assumes idealized conditions: perfect phase matching ($\Delta k = 0$) and negligible absorption ($\gamma_1 = \gamma_2 = \gamma_3 = 0$), which allows us to focus on the depleted-pump dynamics without thermal or phase-mismatch complications. These idealizations are appropriate for demonstrating the toolkit's capabilities in a well-controlled scenario, and the methodology can be extended to include thermal effects and phase mismatch as needed.

The numerical algorithm proceeds as follows: (1) The computational domain is discretized into a three-dimensional grid with $N_t$ time steps, $N_r$ radial steps, and $N_z$ longitudinal steps. A non-uniform radial grid is employed, with finer spacing near the beam axis ($r \leq 5\omega_f$) and coarser spacing in the outer region to optimize computational efficiency while maintaining accuracy in the beam interaction zone. (2) Initial conditions are set at $t=0$ for all spatial points, with the fundamental beams following the Gaussian profiles given in Eqs.~\eqref{eq:19}--\eqref{eq:20} and the second harmonic field initialized to zero. (3) The solution advances through the crystal using a marching algorithm: for each longitudinal position $z$, the algorithm solves the coupled system for all time steps, and for each time step, it solves for all radial positions. At each grid point, the temporal evolution uses backward differences, the radial derivatives use central differences (both first and second order), and the longitudinal propagation proceeds forward along the crystal axis. (4) The nonlinear coupling terms are evaluated at each grid point using the current field values from the three coupled equations. (5) After completing all time steps at a given $z$ position, the computed field values serve as initial conditions for the next longitudinal step. The process continues until the full crystal length is traversed. The complete source code, including detailed implementation notes, convergence criteria, and execution instructions, is available in our GitHub organization repository \href{https://github.com/Second-Harmonic-Generation}{Second Harmonic Generation (SHG)}, enabling full reproducibility of the results presented here.

The optical parameters of the crystal and the relevant physical constants are listed in Table~\ref{tab:1} and Table~\ref{tab:2}, respectively.

\begin{table}[ht]
    \centering
    \caption{The optical crystal parameters of KTP}
    \label{tab:1}
    \vskip .1in
    \begin{tabular}{|c|c|c|}\hline
        Crystal length                  & $L_c=2$ cm                                  & \cite{seidel1997numerical_21}       \\ \hline
        Radius                          & r = 2 mm                                       & \cite{seidel1997numerical_21}       \\ \hline
        Effective nonlinear coefficient & $d_{eff}=7.3 \,\mathrm{pm/V}$                     & \cite{sabaeian2010investigation_17} \\ \hline
        Ordinary refractive index       & $n^{o, \omega}=1.8296$                       & \cite{kato1991parametric_22}        \\ \hline
        Extraordinary refractive index  & $n^{\mathrm{e}, \omega}=1.7466$              & \cite{kato1991parametric_22}        \\ \hline
        Extraordinary refractive index  & $n^{e, 2 \omega}=1.7881$                     & \cite{kato1991parametric_22}        \\ \hline
        Crystal cutting angles          & $\theta=90^{\circ},\; \varphi=24.77^{\circ}$ & \cite{sabaeian2010investigation_17} \\ \hline
    \end{tabular}
\end{table}

\begin{table}[ht]
    \centering
    \caption{The physical constants used in the coupled field equations}
    \label{tab:2}
    \vskip .1in
    \begin{tabular}{|c|c|}\hline
        Fundamental wavelength                    & $\lambda_{1}=1064 \;nm$ \\ \hline
        Second harmonic wavelength                & $\lambda_{2}=532 \;nm$  \\ \hline
        Pulse duration                            & $t_{p}=50 \;\mu s$      \\ \hline
        Pulse energy                              & E = 0.45 J              \\ \hline
        Beam spot size                            & $\omega_{f}=80 \;\mu m$ \\ \hline
        Number of time steps                      & Nt = 2511               \\ \hline
        Number of steps in radial direction       & Nr = 120                \\ \hline
        Number of steps in longitudinal direction & Nz = 12000              \\ \hline
        Absorption coefficient (ideal case)       & $\gamma_1 = \gamma_2 = \gamma_3 = 0$ \\ \hline
        Absorption coefficient (with absorption)  & $\gamma_1 = \gamma_2 = 0.5 \,\mathrm{m}^{-1}$, $\gamma_3 = 4 \,\mathrm{m}^{-1}$ \\ \hline
    \end{tabular}
\end{table}

\noindent
The grid parameters (Nt, Nr, Nz) were selected based on convergence studies from our previous numerical work \cite{sabaeian2014heat, Sabaeian2026CWGCoupled, sabaeian2014pulsed, sabaeian2026depleted}, where these discretization choices were shown to provide converged solutions for similar SHG configurations.

\begin{figure}[!htbp]
    \centering
    \begin{subfigure}[t]{0.48\textwidth}
        \centering
        \includegraphics[width=\linewidth]{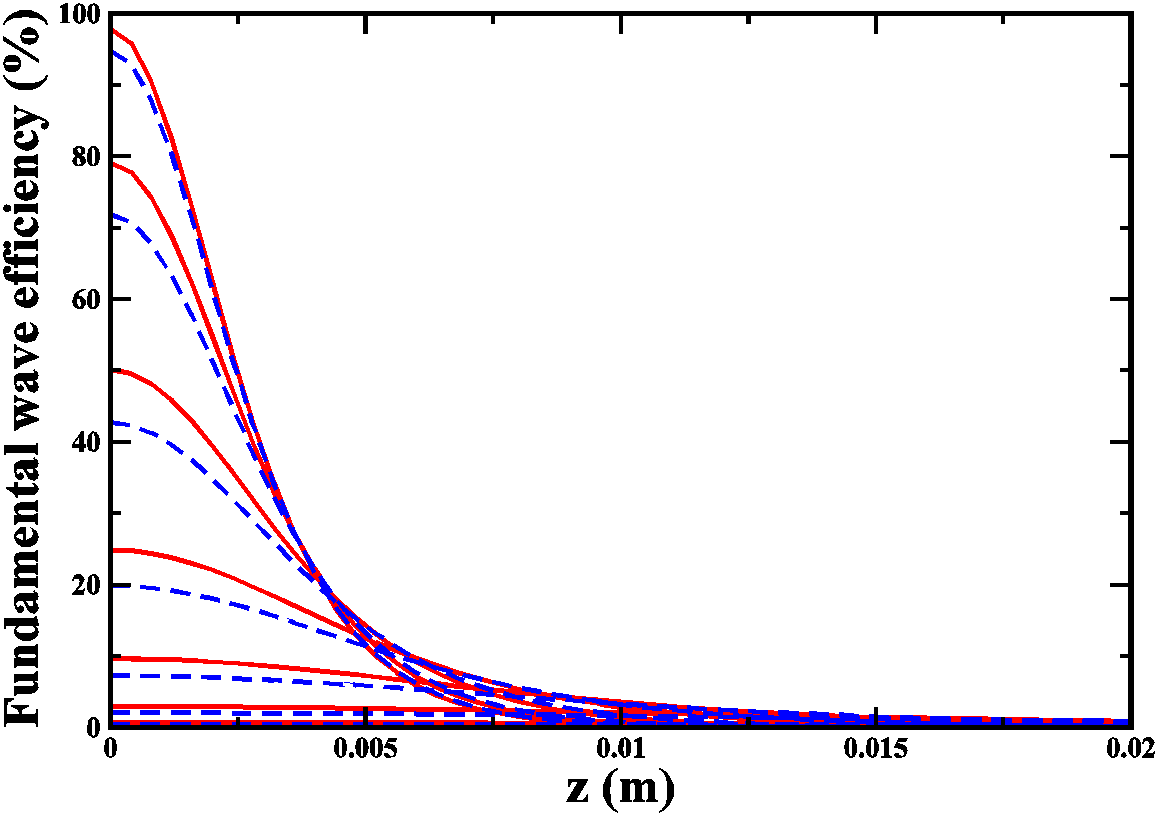}
        \caption{Efficiency of the fundamental wave along the crystal axis ($r = 0$) for several times: from the top to the bottom solid curves are $t= 25\,\mu\mathrm{s}, 75\,\mu\mathrm{s}, 100\,\mu\mathrm{s}$, and from the top to the bottom dashed curves are $t=125\,\mu\mathrm{s}, 150\,\mu\mathrm{s}, 175\,\mu\mathrm{s}, 200\,\mu\mathrm{s}$.}
        \label{fig:1}
    \end{subfigure}
    \hfill
    \begin{subfigure}[t]{0.48\textwidth}
        \centering
        \includegraphics[width=\linewidth]{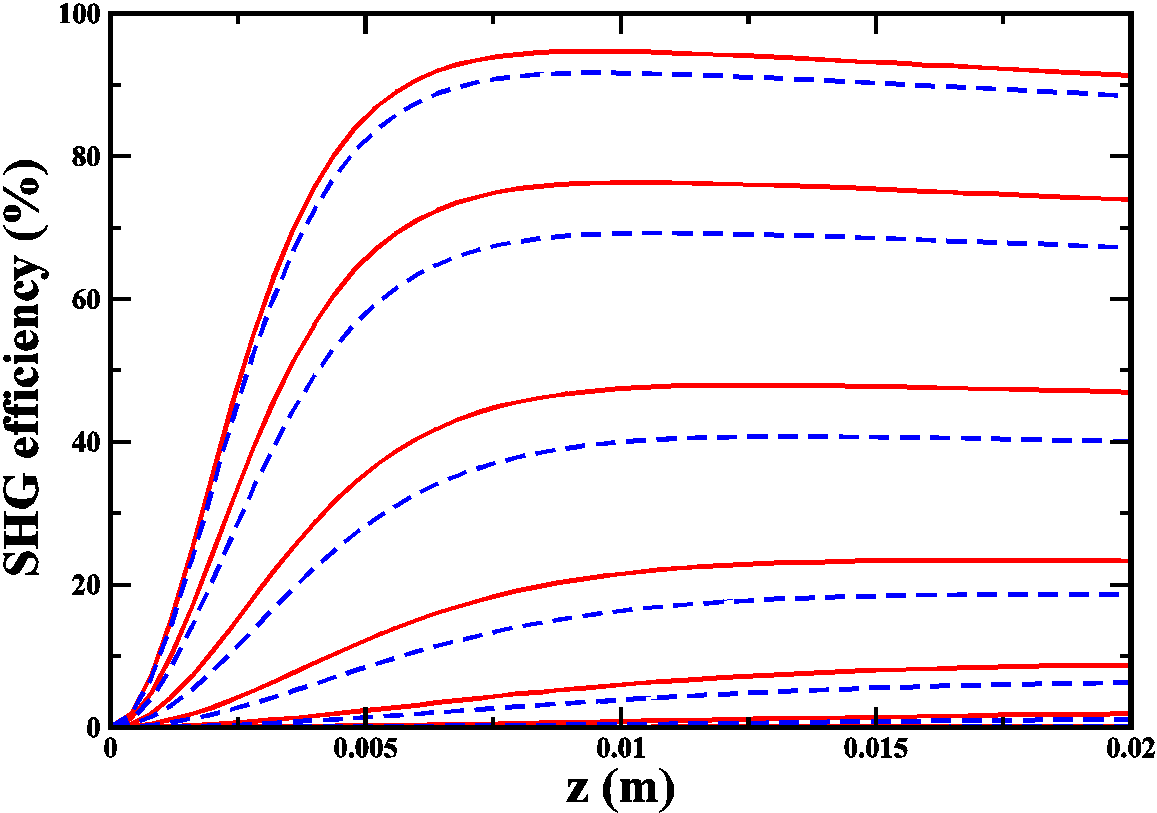}
        \caption{Efficiency of the second harmonic generation along the crystal length ($r = 0$) at several times as used in Figure~\ref{fig:1}.}
        \label{fig:2}
    \end{subfigure}
\end{figure}

Figure~\ref{fig:1} shows the efficiency of the fundamental wave (FW) along the crystal axis for different times from $t = 0$ to $t = 200 \,\mu\mathrm{s}$. The solid (red) curves correspond to times between $25 \,\mu\mathrm{s}$ and $100 \,\mu\mathrm{s}$, with efficiency near zero at $t = 25 \,\mu\mathrm{s}$ (when the pulse has not yet fully arrived) and reaching 100 percent at $t = 100 \,\mu\mathrm{s}$, and the dashed (blue) curves correspond to times from $125 \,\mu\mathrm{s}$ to $200 \,\mu\mathrm{s}$, beginning from almost 90 percent and reaching zero at $t = 200 \,\mu\mathrm{s}$. Note that $t_p = 50 \,\mu\mathrm{s}$ is the pulse duration (full width at $1/e$ of maximum), so $t = 100 \,\mu\mathrm{s} = 2 t_p$ corresponds to the time when the pulse envelope reaches its peak, and $t = 200 \,\mu\mathrm{s} = 4 t_p$ represents a time well after the pulse has decayed. For each time, the FW efficiency drops steadily to zero, indicating that the FW transfers its energy to the second harmonic wave (SHW), consistent with depleted-pump behavior; this behavior is also evident in Figure~\ref{fig:2}. In that case, the SHW efficiencies are shown as a function of time. The solid red curves indicate an increase in the SHW field, while the dashed blue curves represent its decrease. According to this figure, under the idealized conditions described in the methodology section, the model predicts near-complete conversion at $t = 2 t_p$. After propagating approximately $z = 5 \,\mathrm{mm}$ through the crystal, the efficiency decreases slightly because of SHW absorption in the crystal.

\begin{figure}[!htbp]
    \centering
    \begin{subfigure}[t]{0.48\textwidth}
        \centering
        \includegraphics[width=\linewidth]{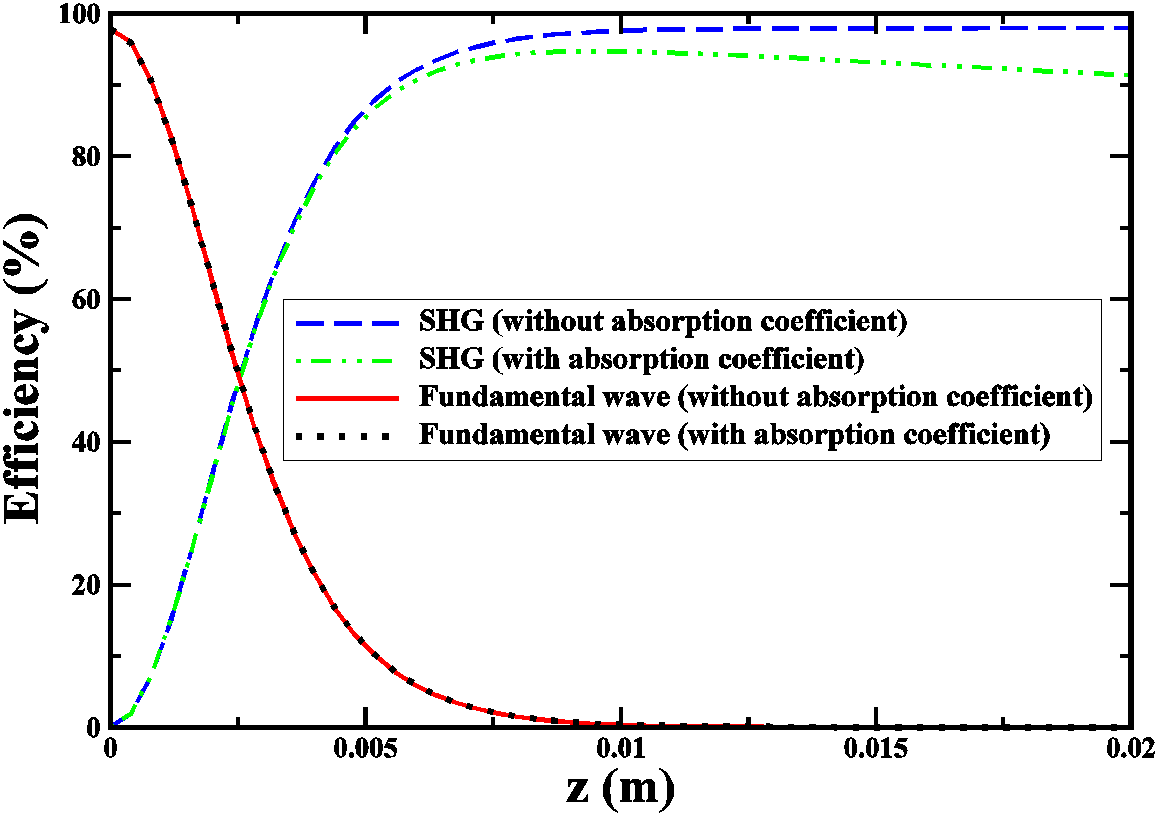}
        \caption{Efficiency of the second harmonic generation and the fundamental wave along the crystal length at $r = 0$ and $t = 2 t_p = 100 \,\mu\mathrm{s}$ (pulse peak).}
        \label{fig:3}
    \end{subfigure}
    \hfill
    \begin{subfigure}[t]{0.48\textwidth}
        \centering
        \includegraphics[width=\linewidth]{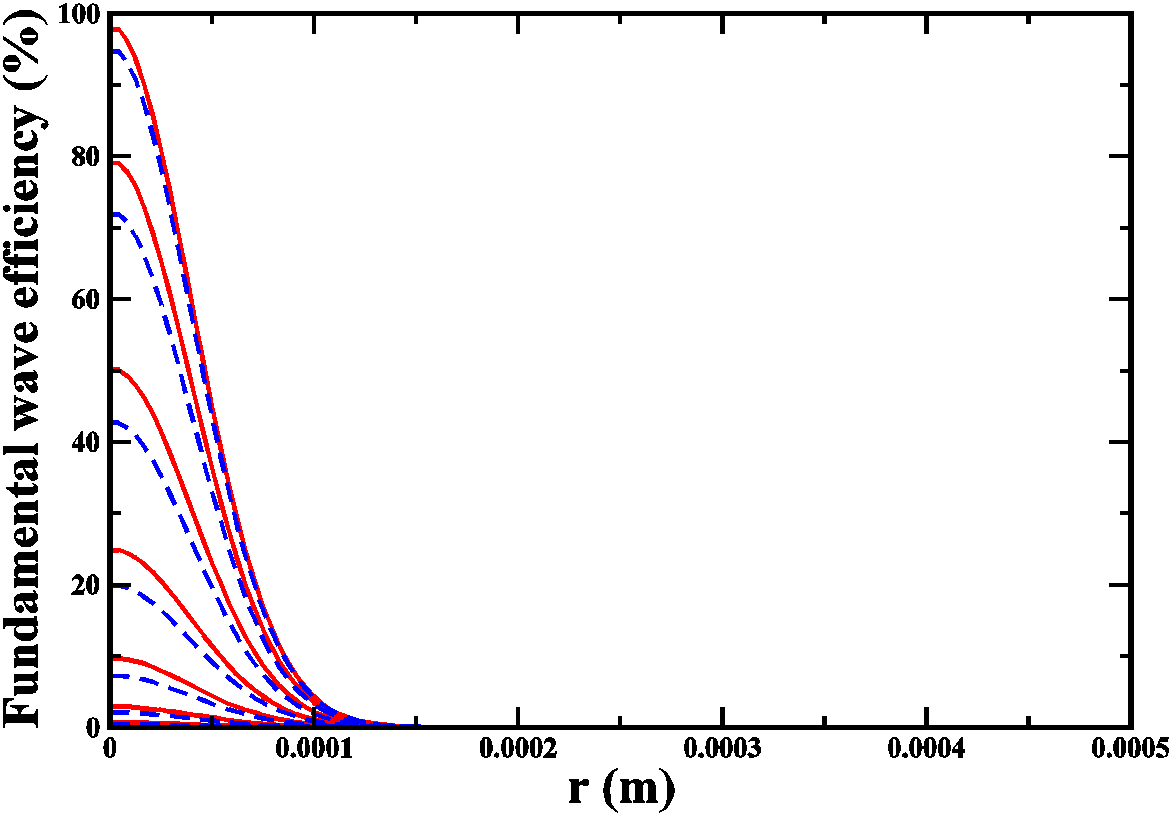}
        \caption{Efficiency of the fundamental wave in the radial direction at the input surface of the crystal ($z = 0$). Various curves correspond to the times shown in Figure~\ref{fig:1}: $t= 25\,\mu\mathrm{s}, 75\,\mu\mathrm{s}, 100\,\mu\mathrm{s}, 125\,\mu\mathrm{s}, 150\,\mu\mathrm{s}, 175\,\mu\mathrm{s}, 200\,\mu\mathrm{s}$.}
        \label{fig:4}
    \end{subfigure}
\end{figure}

Figure~\ref{fig:3} displays the variations of the FW and SHW along the crystal axis at $t = 2 t_p$, when the pulse reaches its maximum energy. Two different cases are compared: the case without optical absorption ($\gamma_1 = \gamma_2 = \gamma_3 = 0$), shown by the dotted curve for the FW and the dash-dotted curve for the SHW efficiencies, and the case with optical absorption ($\gamma_1 = \gamma_2 = 0.5 \,\mathrm{m}^{-1}$, $\gamma_3 = 4 \,\mathrm{m}^{-1}$), shown by the solid curve for the FW and the dashed curve for the SHW efficiencies. As shown, the energy conversion between FW and SHW occurs over a distance of about $5 \,\mathrm{mm}$, where nearly 90\% of FW energy is converted. Since this conversion distance ($5 \,\mathrm{mm}$) is short compared to the crystal length ($L_c = 2 \,\mathrm{cm}$), a depleted formalism is required, and the constant-beam approximation for the FW is no longer valid. Figure~\ref{fig:4} shows the radial profile of the FW efficiency at the entrance surface of the crystal ($z = 0$). The incident FW on the crystal face has a Gaussian profile. Before the onset of energy exchange, the FW efficiency at the center of the entrance face is 100 percent and decreases gradually toward the lateral surface of the crystal.

\begin{figure}[!htbp]
    \centering
    \begin{subfigure}[t]{0.48\textwidth}
        \centering
        \includegraphics[width=\linewidth]{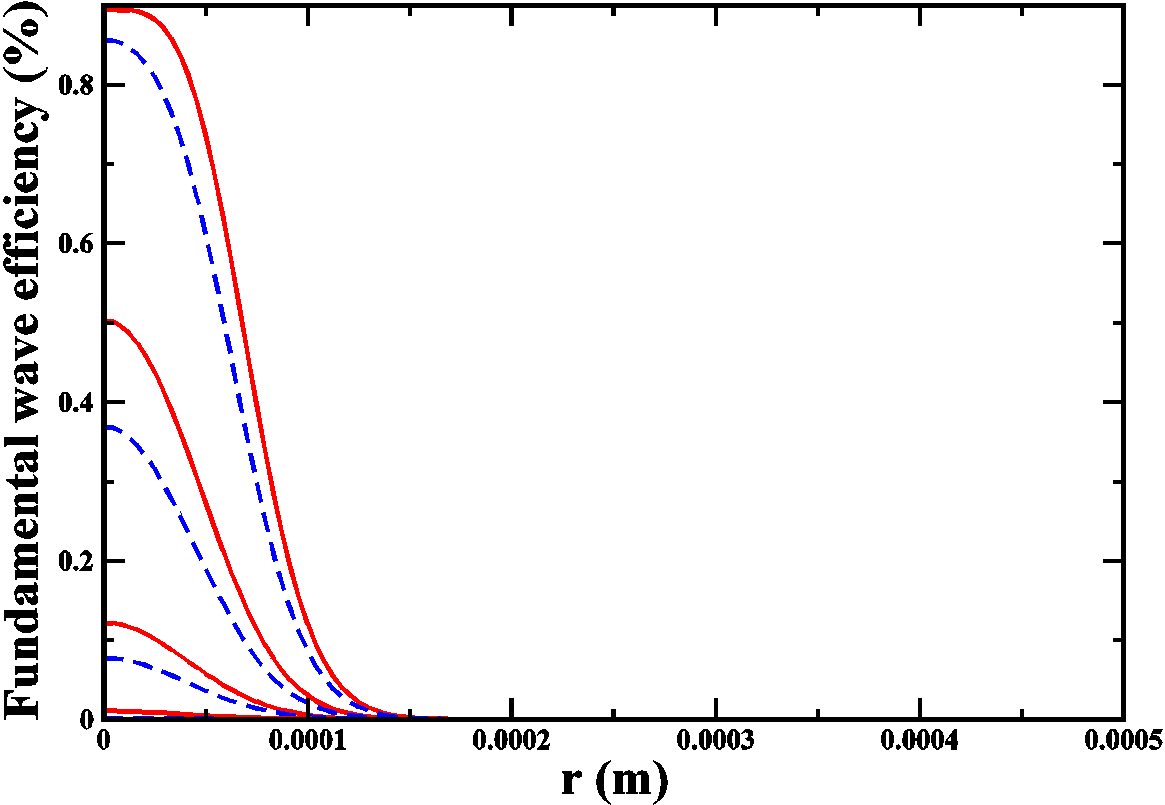}
        \caption{Efficiency of the fundamental wave in the radial direction at the output surface of the crystal ($z = 2 \,\mathrm{cm}$).}
        \label{fig:5}
    \end{subfigure}
    \hfill
    \begin{subfigure}[t]{0.48\textwidth}
        \centering
        \includegraphics[width=\linewidth]{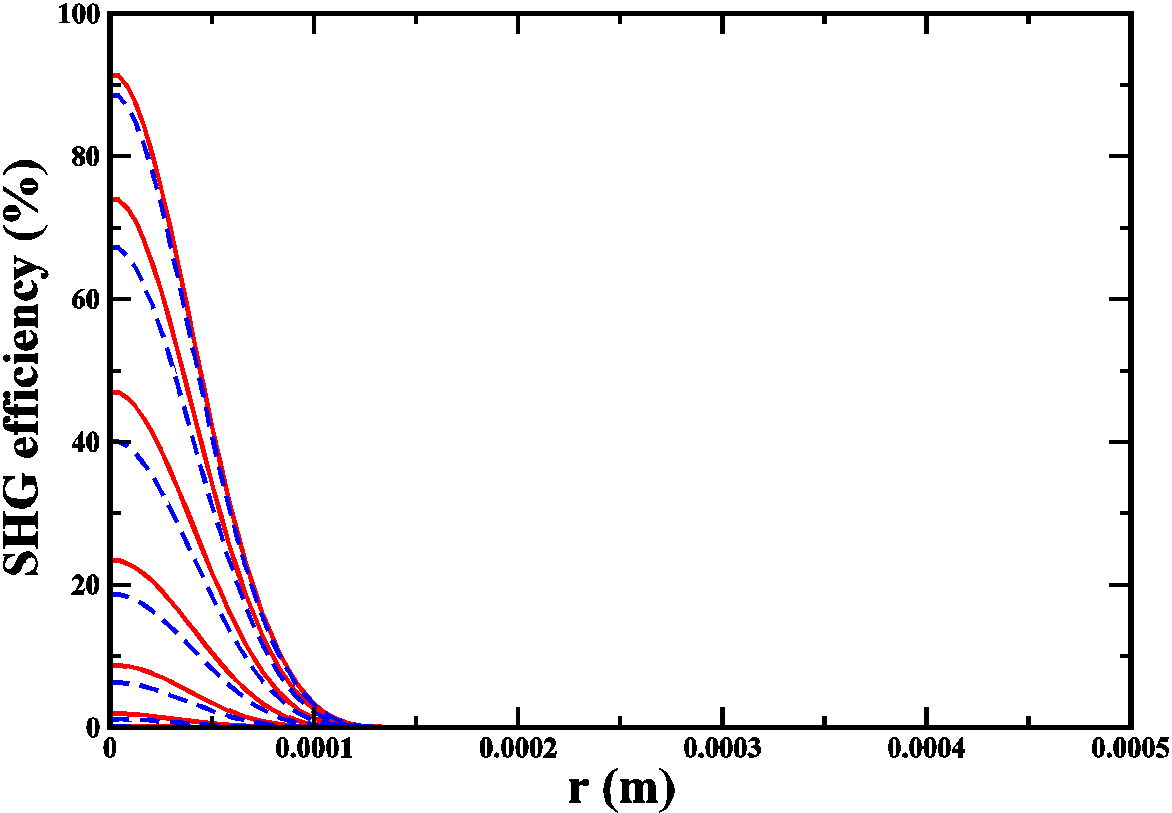}
        \caption{Efficiency of the SHG in the radial direction at the output surface of the crystal ($z = 2 \,\mathrm{cm}$).}
        \label{fig:6}
    \end{subfigure}
\end{figure}

Figure~\ref{fig:5} shows that at the exit face the FW efficiency has decreased to approximately 0.5\%. At the same time, a Gaussian SHW profile is generated in the crystal as a result of the energy conversion. Figure~\ref{fig:6} shows the transverse Gaussian profile of the SHW at the output surface of the crystal at $z = 2 \, \mathrm{cm}$. As expected for an ideal depleted mechanism, the SHW efficiency at the exit face is very high, and the SHW retains the same transverse Gaussian profile.

\begin{figure}[!htbp]
    \centering
    \begin{subfigure}[t]{0.48\textwidth}
        \centering
        \includegraphics[width=\linewidth]{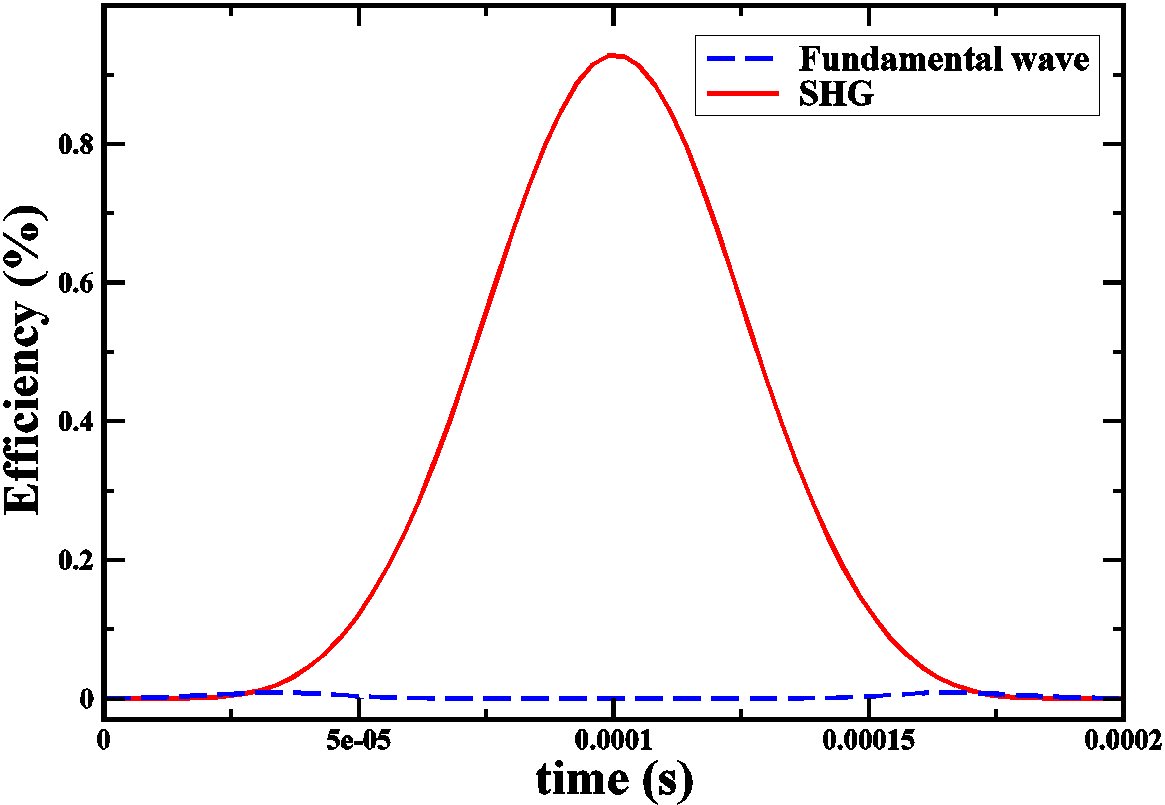}
        \caption{The efficiency of the fundamental wave (dashed curve) and second harmonic generation (solid curve) at the output face of the crystal ($z = 2 \,\mathrm{cm}$, $r = 0$) from $t=0$ to $t=4 t_p = 200 \,\mu\mathrm{s}$.}
        \label{fig:7}
    \end{subfigure}
    \hfill
    \begin{subfigure}[t]{0.48\textwidth}
        \centering
        \includegraphics[width=\linewidth]{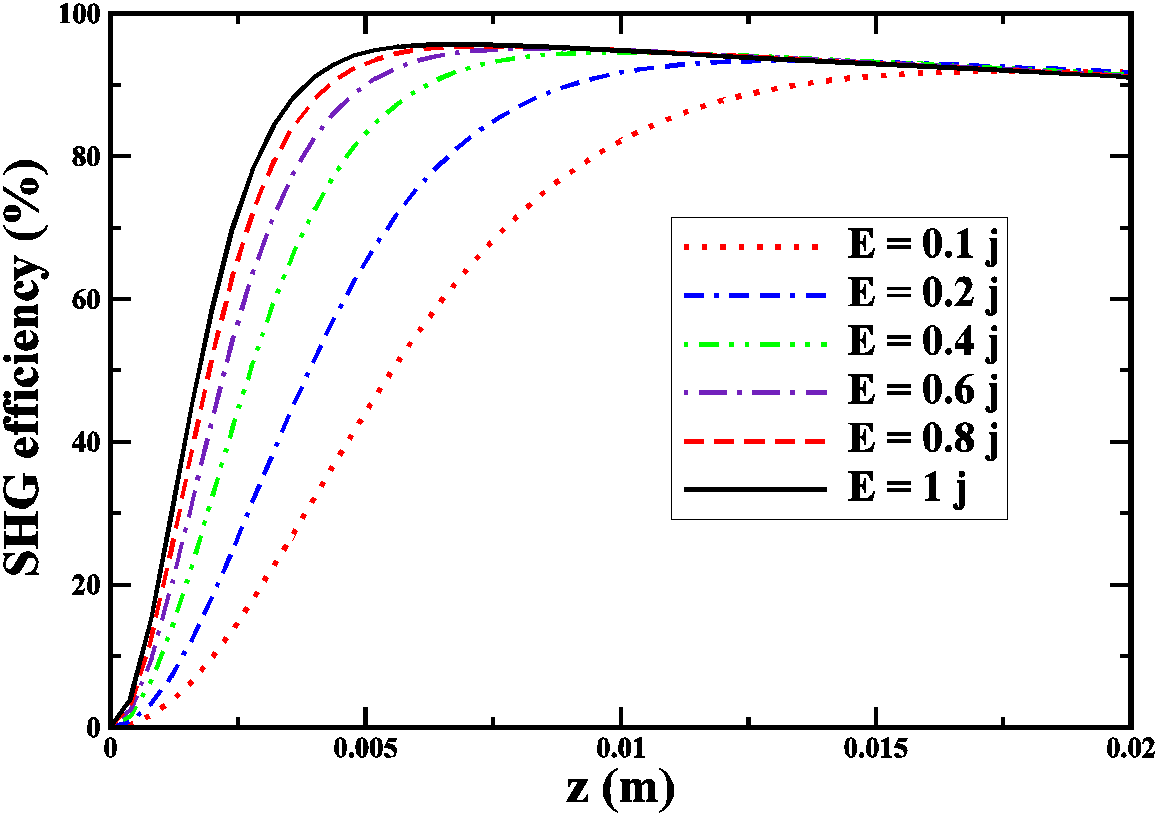}
        \caption{Efficiency of the second harmonic generation along the crystal length for different pulse energies: $0.1 \,\mathrm{J}, 0.2 \,\mathrm{J}, 0.4 \,\mathrm{J}, 0.6 \,\mathrm{J}, 0.8 \,\mathrm{J}, 1.0 \,\mathrm{J}$.}
        \label{fig:8}
    \end{subfigure}
\end{figure}

Figure~\ref{fig:7} shows the temporal variation of the FW and SHW efficiencies at the central point ($r = 0$) of the output surface of the crystal. Under the idealized conditions described in the methodology section, the FW energy (dashed curve) is completely converted into SHW energy (solid curve). The variations of SHW efficiency are further investigated by changing the pulse energy, using $0.1 J,\;0.2 J,\;0.4 J,\;0.6 J,\;0.8 J,$ and $1 J$, as shown in Figure~\ref{fig:8}. As illustrated in Figure~\ref{fig:8} and consistent with Eq.~\eqref{eq:17}, higher energies lead to a shorter interaction length and thus faster energy conversion. More precisely, for higher energies the conversion of energy between FW and SHW occurs at shorter distances from the input surface.

\section{Conclusion}
\noindent
This case study demonstrates the toolkit suite's extensibility by showing its modular architecture, validated implementations, and accessibility. We introduced a three-dimensional, spatiotemporal model for pulsed Gaussian second harmonic waves (SHW) based on three coupled equations describing Type II SHG. Despite the complexity and coupled nature of these equations, we employed numerical methods to obtain solutions and showed how a computational toolkit can be extended within our suite. 

Future studies could investigate other wave types (e.g., Bessel--Gaussian waves) or alternative crystal configurations. Experimental validation, based on comparing measured results with numerical predictions, would further improve the predictive power and practical relevance of the model. It could also inform strategies for controlling crystal properties and designing new nonlinear optical configurations. We are currently extending and applying the toolkit suite to additional SHG problems, and we will report the corresponding results in separate studies.

Our methodology is adaptable and can be used to model complex, challenging nonlinear wave-related phenomena within the same computational environment. By making the codes publicly available and well-documented, the SHG Computational Toolkit Suite provides a reusable foundation for adapting individual toolkits, developing new ones for additional scenarios, and reusing existing numerical components without reconstructing the computational infrastructure from scratch. 

M. M. Rezaee, M. Sabaeian, A. Motazedian, F. S. Jalil-Abadi, and M. Ghadri would like to thank Shahid Chamran University of Ahvaz for supporting this work.

\section{Disclosures} 
\noindent
The authors declare no conflicts of interest.

%%%%%%%%%%%%%%%%%%%%%%% References %%%%%%%%%%%%%%%%%%%%%%%%%
\bibliography{references}

@article{sabaeian2010investigation_17,
  title={Investigation of thermally-induced phase mismatching in continuous-wave second harmonic generation: a theoretical model},
  author={Sabaeian, Mohammad and Mousave, Laleh and Nadgaran, Hamid},
  journal={Optics express},
  volume={18},
  number={18},
  pages={18732--18743},
  year={2010},
  publisher={Optical Society of America}
}

@article{sabaeian2008analytical,
  title     = {Analytical solution of the heat equation in a longitudinally pumped cubic solid-state laser},
  author    = {Sabaeian, Mohammad and Nadgaran, Hamid and Mousave, Laleh},
  journal   = {Applied optics},
  volume    = {47},
  number    = {13},
  pages     = {2317--2325},
  year      = {2008},
  publisher = {Optical Society of America}
}

@article{sabaeian2015tempdist,
  title     = {Temperature distribution in a Gaussian end-pumped nonlinear KTP crystal: the temperature dependence of thermal conductivity and radiation boundary condition},
  author    = {Sabaeian, Mohammad and Jalil-Abadi, Fatemeh Sedaghat and Rezaee, Mostafa Mohammad and Motazedian, Alireza and Shahzadeh, Mohammadreza},
  journal   = {Brazilian journal of physics},
  volume    = {45},
  number    = {1},
  pages     = {1--9},
  year      = {2015},
  publisher = {Springer}
}

@article{rezaee2015complete,
  title     = {Complete anisotropic time-dependent heat equation in KTP crystal under repetitively pulsed Gaussian beams: a numerical approach},
  author    = {Rezaee, Mostafa Mohammad and Sabaeian, Mohammad and Motazedian, Alireza and Jalil-Abadi, Fatemeh Sedaghat and Khaldi-Nasab, Ali},
  journal   = {Applied Optics},
  volume    = {54},
  number    = {6},
  pages     = {1241--1249},
  year      = {2015},
  publisher = {Optica Publishing Group}
}

@article{rezaee2015thermally,
  title     = {Thermally induced phase mismatching in a repetitively Gaussian pulsed pumping KTP crystal: a spatiotemporal treatment},
  author    = {Rezaee, Mostafa Mohammad and Sabaeian, Mohammad and Motazedian, Alireza and Jalil-Abadi, Fatemeh Sedaghat and Askari, Hadi and Khazrk, Iman},
  journal   = {Applied Optics},
  volume    = {54},
  number    = {15},
  pages     = {4781--4788},
  year      = {2015},
  publisher = {Optica Publishing Group}
}

@article{sabaeian2015temperature,
  title     = {Temperature increase effects on a double-pass cavity type II second-harmonic generation: a model for depleted Gaussian continuous waves},
  author    = {Sabaeian, Mohammad and Jalil-Abadi, Fatemeh Sedaghat and Rezaee, Mostafa Mohammad and Motazedian, Alireza and Shahzadeh, Mohammadreza},
  journal   = {Applied optics},
  volume    = {54},
  number    = {4},
  pages     = {869--875},
  year      = {2015},
  publisher = {Optica Publishing Group}
}

@article{sabaeian2014pulsed,
  title     = {Pulsed Bessel--Gauss beams: a depleted wave model for type II second-harmonic generation},
  author    = {Sabaeian, Mohammad and Motazedian, Alireza and Rezaee, Mostafa Mohammad and Jalil-Abadi, Fatemeh Sedaghat},
  journal   = {Applied Optics},
  volume    = {53},
  number    = {32},
  pages     = {7691--7696},
  year      = {2014},
  publisher = {Optica Publishing Group}
}

@article{sabaeian2014heat,
  title     = {Heat coupled Gaussian continuous-wave double-pass type-II second harmonic generation: inclusion of thermally induced phase mismatching and thermal lensing},
  author    = {Sabaeian, Mohammad and Jalil-Abadi, Fatemeh Sedaghat and Rezaee, Mostafa Mohammad and Motazedian, Alireza},
  journal   = {Optics express},
  volume    = {22},
  number    = {21},
  pages     = {25615--25628},
  year      = {2014},
  publisher = {Optica Publishing Group}
}

@misc{Rezaee2025ThermalKTP,
      title={A Thermal Modeling Toolkit for Continuous-Wave Gaussian Second-Harmonic Generation in KTP Crystal}, 
      author={Mostafa M. Rezaee and Mohammad Sabaeian and Alireza Motazedian and Fatemeh Sedaghat Jalil-Abadi and Mohammad Ghadri},
      year={2025},
      eprint={2512.12145},
      archivePrefix={arXiv},
      primaryClass={physics.optics},
      url={https://arxiv.org/abs/2512.12145}, 
      DOI={https://doi.org/10.48550/arXiv.2512.12145},
      note={DOI: 10.48550/arXiv.2512.12145}
}

@article{Rezaee_2026_CWG_Heat,
title={A Toolkit for Time-Dependent 3D Thermal Modeling in KTP Crystal under Pulsed-Gaussian Second-Harmonic Generation Operation},
url={http://dx.doi.org/10.36227/techrxiv.176972410.05925345/v1},
DOI={10.36227/techrxiv.176972410.05925345/v1},
journal = {TechRxiv},
publisher={Institute of Electrical and Electronics Engineers (IEEE)},
author={Rezaee, Mostafa M. and Sabaeian, Mohammad and Motazedian, Alireza and Jalil-Abadi, Fatemeh Sedaghat and Ghadri, Mohammad},
year={2026},
month=jan,
note={DOI: 10.36227/techrxiv.176972410.05925345/v1}
}

@article{Rezaee_2026_PWG-PhM,
title={Temperature-Dependent Phase Mismatch in KTP Crystal: An Open-Source Computational Model},
url={http://dx.doi.org/10.36227/techrxiv.177130658.80386246/v1},
DOI={10.36227/techrxiv.177130658.80386246/v1},
journal = {TechRxiv},
publisher={Institute of Electrical and Electronics Engineers (IEEE)},
author={Rezaee, Mostafa M. and Sabaeian, Mohammad and Motazedian, Alireza and Jalil-Abadi, Fatemeh Sedaghat and Khazrak, Iman and Ghadri, Mohammad},
year={2026},
month=feb,
note={DOI: 10.36227/techrxiv.177130658.80386246/v1}
}

@article{Sabaeian_2026_CWG-Ideal,
title={Depleted Gaussian Continuous-Wave Second Harmonic Generation: An Open Source Study on Modeling Electric-Field Distribution and Thermal Effects in a KTP Double-Pass Cavity},
url={http://dx.doi.org/10.36227/techrxiv.177208104.40622819/v1},
DOI={10.36227/techrxiv.177208104.40622819/v1},
journal = {TechRxiv},
publisher={Institute of Electrical and Electronics Engineers (IEEE)},
author={Sabaeian, Mohammad and Jalil-Abadi, Fatemeh Sedaghat and Rezaee, Mostafa M and Motazedian, Alireza and Ghadri, Mohammad},
year={2026},
month=feb,
note={DOI: 10.36227/techrxiv.177208104.40622819/v1}
}

@article{sabaeian2026depleted,
  author  = {Mohammad Sabaeian and Alireza Motazedian and Mostafa M. Rezaee and Fatemeh Sedaghat Jalil-Abadi and Mohammad Ghadri},
  title   = {Depleted-Wave Electric-Field Solver for Pulsed Bessel-Gaussian Type-II Frequency Doubling in KTP Crystals},
  journal = {Research Square},
  year    = {2026},
  month   = mar,
  day     = {25},
  doi     = {10.21203/rs.3.rs-9203079/v1},
  url     = {https://doi.org/10.21203/rs.3.rs-9203079/v1},
}

@article{Sabaeian2026CWGCoupled,
	doi = {10.20944/preprints202604.0962.v1},
	url = {https://doi.org/10.20944/preprints202604.0962.v1},
	year = 2026,
	month = {April},
	publisher = {Preprints},
	author = {Mohammad Sabaeian and Alireza Motazedian and Mostafa M. Rezaee and Fatemeh Sedaghat Jalil-Abadi and Mohammad Ghadri},
	title = {Thermo-Optical Modeling of Double-Pass Type-II Second Harmonic Generation in KTP},
	journal = {Preprints}
}

@article{PhysRevB.104.L161202,
  title = {Current-induced second harmonic generation in inversion-symmetric Dirac and Weyl semimetals},
  author = {Takasan, Kazuaki and Morimoto, Takahiro and Orenstein, Joseph and Moore, Joel E.},
  journal = {Phys. Rev. B},
  volume = {104},
  issue = {16},
  pages = {L161202},
  numpages = {7},
  year = {2021},
  month = {Oct},
  publisher = {American Physical Society},
  doi = {10.1103/PhysRevB.104.L161202},
  url = {https://link.aps.org/doi/10.1103/PhysRevB.104.L161202}
}

@article{Briggs:24,
author = {Ian Briggs and Paokang Chen and Linran Fan},
journal = {Opt. Lett.},
number = {23},
pages = {6637--6640},
publisher = {Optica Publishing Group},
title = {Precise wavelength alignment of second-harmonic generation in thin-film lithium niobate resonators},
volume = {49},
month = {Dec},
year = {2024},
url = {https://opg.optica.org/ol/abstract.cfm?URI=ol-49-23-6637},
doi = {10.1364/OL.540614}
}

@Article{nano14080662,
AUTHOR = {Fu, Yue and Liu, Zhengyan and Yue, Song and Zhang, Kunpeng and Wang, Ran and Zhang, Zichen},
TITLE = {Optical Second Harmonic Generation of Low-Dimensional Semiconductor Materials},
JOURNAL = {Nanomaterials},
VOLUME = {14},
YEAR = {2024},
NUMBER = {8},
ARTICLE-NUMBER = {662},
URL = {https://www.mdpi.com/2079-4991/14/8/662},
PubMedID = {38668156},
ISSN = {2079-4991},
DOI = {10.3390/nano14080662}
}

@article{aghigh2023second,
  title={Second harmonic generation microscopy: a powerful tool for bio-imaging},
  author={Aghigh, Arash and Bancelin, St{\'e}phane and Rivard, Maxime and Pinsard, Maxime and Ibrahim, Heide and L{\'e}gar{\'e}, Fran{\c{c}}ois},
  journal={Biophysical Reviews},
  volume={15},
  number={1},
  pages={43--70},
  year={2023},
  publisher={Springer}
}

@article{zu2024optical,
  title={Optical second harmonic generation in anisotropic multilayers with complete multireflection of linear and nonlinear waves using \#SHAARP. ml package},
  author={Zu, Rui and Wang, Bo and He, Jingyang and Weber, Lincoln and Saha, Akash and Chen, Long-Qing and Gopalan, Venkatraman},
  journal={npj Computational Materials},
  volume={10},
  number={1},
  pages={64},
  year={2024},
  publisher={Nature Publishing Group UK London}
}

@article{liu2019phase,
  title={Phase-matched second-harmonic generation in coupled nonlinear optical waveguides},
  author={Liu, Bodong and Yu, Huakang and Li, Zhi-yuan and Tong, Limin},
  journal={Journal of the Optical Society of America B},
  volume={36},
  number={10},
  pages={2650--2658},
  year={2019},
  publisher={Optical Society of America}
}

@article{AKRAMOV2024129827,
title = {Second-harmonic generation in branched optical waveguides: Metric graphs based approach},
journal = {Physics Letters A},
volume = {524},
pages = {129827},
year = {2024},
issn = {0375-9601},
doi = {https://doi.org/10.1016/j.physleta.2024.129827},
url = {https://www.sciencedirect.com/science/article/pii/S0375960124005206},
author = {M. Akramov and B. Eshchanov and S. Usanov and Sh. Norbekov and D. Matrasulov},
}

@article{zu2022analytical,
  title={Analytical and numerical modeling of optical second harmonic generation in anisotropic crystals using \#SHAARP package},
  author={Zu, Rui and Wang, Bo and He, Jingyang and Wang, Jian-Jun and Weber, Lincoln and Chen, Long-Qing and Gopalan, Venkatraman},
  journal={npj Computational Materials},
  volume={8},
  number={1},
  pages={246},
  year={2022},
  publisher={Nature Publishing Group UK London}
}

@article{boudjema2020pyshs,
  title={PySHS: Python Open Source Software for Second Harmonic Scattering},
  author={Boudjema, Lotfi and Aarrass, Hanna and Assaf, Marwa and Morille, Marie and Martin-Gassin, Gaelle and Gassin, Pierre-Marie},
  journal={Journal of Chemical Information and Modeling},
  volume={60},
  number={12},
  pages={5912--5917},
  year={2020},
  publisher={ACS Publications}
}

@article{paez2024cuda,
  title={CUDA-based focused Gaussian beams second-harmonic generation efficiency calculator},
  author={P{\'a}ez-L{\'o}pez, R. and Rama{\'\i}rez, J. F. and Garc{\'\i}a, H. and Carrascosa, M.},
  journal={Computer Physics Communications},
  volume={301},
  pages={109238},
  year={2024},
  publisher={Elsevier},
  doi={10.1016/j.cpc.2024.109238}
}

@article{Zhao:20,
author = {Chuanrui Zhao and Xinle Wang and Zhengping Wang and Yuxiang Sun and Shiwei Tian and Hongkai Ren and Fapeng Yu and Xian Zhao and Xinguang Xu},
journal = {Opt. Express},
number = {22},
pages = {33274--33284},
publisher = {Optica Publishing Group},
title = {Remarkable temperature-dependent second-harmonic-generation performance of a YCOB crystal},
volume = {28},
month = {Oct},
year = {2020},
url = {https://opg.optica.org/oe/abstract.cfm?URI=oe-28-22-33274},
doi = {10.1364/OE.410606}
}

@article{li2024highly,
  title={Highly efficient second-harmonic generation in a double-layer thin-film lithium niobate waveguide},
  author={Li, Yuan and Zhang, Xiuquan and Cai, Lutong and Zhang, Lin},
  journal={Advanced Photonics Nexus},
  volume={3},
  number={6},
  pages={066009--066009},
  year={2024},
  publisher={Society of Photo-Optical Instrumentation Engineers}
}

@article{Szarvas:18,
author = {Tamas Szarvas and Zsolt Kis},
journal = {J. Opt. Soc. Am. B},
number = {4},
pages = {731--740},
publisher = {Optica Publishing Group},
title = {Numerical simulation of nonlinear second harmonic wave generation by the finite difference frequency domain method},
volume = {35},
month = {Apr},
year = {2018},
url = {https://opg.optica.org/josab/abstract.cfm?URI=josab-35-4-731},
doi = {10.1364/JOSAB.35.000731},
}

@article{Carnio:20,
author = {B. N. Carnio and A. Y. Elezzabi},
journal = {Opt. Lett.},
number = {13},
pages = {3657--3660},
publisher = {Optica Publishing Group},
title = {Backward terahertz difference frequency generation via modal phase-matching in a planar LiNbO3 waveguide},
volume = {45},
month = {Jul},
year = {2020},
url = {https://opg.optica.org/ol/abstract.cfm?URI=ol-45-13-3657},
doi = {10.1364/OL.393283},
}

@article{saito2016numerical,
  title={Numerical analysis of second harmonic generation for THz-wave in a photonic crystal waveguide using a nonlinear FDTD algorithm},
  author={Saito, Kyosuke and Tanabe, Tadao and Oyama, Yutaka},
  journal={Optics Communications},
  volume={365},
  pages={164--167},
  year={2016},
  publisher={Elsevier}
}

@article{Chou:98,
author = {Hsu-Feng Chou and Ching-Fuh Lin and Gin-Chung Wang},
journal = {J. Lightwave Technol.},
number = {9},
pages = {1686},
publisher = {Optica Publishing Group},
title = {An Iterative Finite Difference Beam Propagation Method for Modeling Second-Order Nonlinear Effects in Optical Waveguides},
volume = {16},
month = {Sep},
year = {1998},
url = {https://opg.optica.org/jlt/abstract.cfm?URI=jlt-16-9-1686},
}

@article{van2024encore,
  title={ENCORE: a practical implementation to improve reproducibility and transparency of computational research},
  author={van Kampen, Antoine HC and Mahamune, Utkarsh and Jongejan, Aldo and van Schaik, Barbera DC and Balashova, Daria and Lashgari, Danial and Pras-Raves, Mia and Wever, Eric JM and Dane, Adrie D and Garc{\'\i}a-Valiente, Rodrigo and others},
  journal={Nature Communications},
  volume={15},
  number={1},
  pages={8117},
  year={2024},
  publisher={Nature Publishing Group UK London}
}

@article{dogucu2024reproducibility,
  title={Reproducibility in the Classroom},
  author={Dogucu, Mine},
  journal={Annual Review of Statistics and Its Application},
  volume={12},
  year={2024},
  publisher={Annual Reviews}
}

@inproceedings{fredly2024computational,
  title={How Computational Physics Students Improve their Computational Literacy},
  author={Fredly, Karl Henrik and Odden, Tor Ole Bigton and Zwickl, Ben},
  booktitle={Physics Education Research Conference Proceedings},
  pages={138--143},
  year={2024}
}

@article{odden2023using,
  title={Using computational essays to foster disciplinary epistemic agency in undergraduate science},
  author={Odden, Tor Ole B and Silvia, Devin W and Malthe-S{\o}renssen, Anders},
  journal={Journal of Research in Science Teaching},
  volume={60},
  number={5},
  pages={937--977},
  year={2023},
  publisher={Wiley Online Library}
}

@article{weller2022development,
  title={Development and illustration of a framework for computational thinking practices in introductory physics},
  author={Weller, Daniel P and Bott, Theodore E and Caballero, Marcos D and Irving, Paul W},
  journal={Physical Review Physics Education Research},
  volume={18},
  number={2},
  pages={020106},
  year={2022},
  publisher={APS}
}

@article{phillips2023physicality,
  title={Physicality, modeling, and agency in a computational physics class},
  author={Phillips, AM and Gouvea, EJ and Gravel, BE and Beachemin, P-H and Atherton, TJ},
  journal={Physical Review Physics Education Research},
  volume={19},
  number={1},
  pages={010121},
  year={2023},
  publisher={APS}
}

@article{boyd1968parametric,
  title={Parametric interaction of focused Gaussian light beams},
  author={Boyd, GD and Kleinman, DA},
  journal={Journal of Applied Physics},
  volume={39},
  number={8},
  pages={3597--3639},
  year={1968},
  publisher={American Institute of Physics}
}

@article{Boyd2003Nonlinear_19,
  title={Nonlinear optics.},
  author={Boyd, R. W.},
  journal={Academic Press},
  volume={},
  number={},
  pages={ },
  year={2003},
  publisher={Academic Press}
}

@article{feng2012efficient_20,
  title={Efficient Nd: Y 0.5 Gd 0.5 VO 4-KTiOPO 4 green laser under diode pumping into the emitting level 4 F 3/2},
  author={Feng, DW and Feng, Y and Zhang, GW},
  journal={Laser Physics},
  volume={22},
  number={5},
  pages={888--891},
  year={2012},
  publisher={Springer}
}

@inproceedings{seidel1997numerical_21,
  title={Numerical modeling of thermal effects in nonlinear crystals for high-average-power second harmonic generation},
  author={Seidel, Stefan and Mann, Guido},
  booktitle={Modeling and Simulation of Higher-Power Laser Systems IV},
  volume={2989},
  pages={204--214},
  year={1997},
  organization={International Society for Optics and Photonics}
}

@article{kato1991parametric_22,
  title={Parametric oscillation at 3.2 mu m in KTP pumped at 1.064 mu m},
  author={Kato, K},
  journal={IEEE journal of quantum electronics},
  volume={27},
  number={5},
  pages={1137--1140},
  year={1991},
  publisher={IEEE}
}

\end{document}